\documentclass[aps,prb,preprint,superscriptaddress,showkeys,showpacs]{revtex4}

\usepackage{latexsym} \usepackage{amsmath} \usepackage{bbold} \usepackage{dcolumn}
\usepackage{bm} \usepackage{graphicx} \usepackage{epsfig}
\graphicspath{{./Figures/}}

\newcommand{\adag}[2]{{#1}_{#2}^{\dagger}}

\newcommand{\vbbnoon}[1]{\frac{V^{#1}_{bb}}{\omega_{#1}}}
\newcommand{\deltavoon}[1]{\frac{\Delta V^{#1}}{\omega_{#1}}}

\begin{document}


\title{Real Space Green's Function Approach to RIXS}




\author{J. J. Kas} \affiliation{Dept.\ of Physics, BOX 351560, Univ.\ of
Washington Seattle, WA 98195-1560}

\author{J. J. Rehr} \affiliation{Dept.\ of Physics, BOX 351560, Univ.\ of
Washington Seattle, WA 98195-1560}

\author{J. A. Soininen} \affiliation{Dept.\ of Physics, POB 64, FI-00014 University of Helsinki, Finland}

\author{P. Glatzel} \affiliation
{European Synchrotron Radiation Facility, BP220, F-38043 Grenoble, France}

\date{\today}

\begin{abstract}
We present an \textit{ab initio} theory of core- and valence resonant
inelastic x-ray scattering (RIXS) based on a real-space multiple scattering
Green's function formalism and a quasi-boson model Hamiltonian. Simplifying
assumptions are made which lead to an approximation of the RIXS spectrum
in terms of a convolution of an effective x-ray absorption signal with
the x-ray emission 
signal.  Additional many body corrections are incorporated in terms of
an effective energy dependent spectral function.  Example
calculations of RIXS are found to give qualitative agreement with
experimental data.  Our approach also yields simulations of
lifetime-broadening suppressed XAS, as observed in high energy resolution
fluorescence detection experiment (HERFD).
Finally possible improvements to our approach are  briefly discussed.
\end{abstract}



\maketitle

\section{Introduction}
\label{IntroSect}
Resonant inelastic x-ray scattering (RIXS) is a powerful tool for probing
occupied and unoccupied densities of states at high resolution.
Moreover, the RIXS signal
contains valuable information about the many-body excitations
of a system, e.g., those beyond the primary quasi-particle
excitation.\cite{glatzel2009,degroot2008,kotani2001,schulke2007}
However, because the RIXS signal is described by the Kramers-Heisenberg
equation rather than Fermi's golden rule and is sensitive to
many-body excitations, theoretical
calculations of RIXS are more difficult than those of related
core-level spectroscopies such as x-ray absorption (XAS), x-ray
emission (XES), and electron energy loss (EELS).
Even so, models of the RIXS spectrum based on single particle band structure
arguments can be quite useful for systems with weak electron correlations,
i.e., $sp$-electron
systems.\cite{ma1992,ma1994,carlisle1995,shin1996,johnson1994,kokko2003,luo1993,fujikawa2004} 
In order to account for particle-hole interactions, methods based on the
Bethe-Salpeter equation have been employed.\cite{shirley2001} 
In addition, there have been works modeling $d$-electron systems within
the single particle picture.\cite{jimenez1999,glatzel2010,veenendaal2006}
Also,
for systems with strong electron correlations, the Anderson impurity
model and atomic multiplet theories have been used to explain
RIXS spectra of localized $d-$ and $f$-state systems.\cite{kotani2005} For more detailed
reviews see e.g., Ref.~\onlinecite{degroot2008,kotani2001,schulke2007}. 

In this paper we introduce a theoretical treatment of RIXS based on
an approximation to the Kramers-Heisenberg equation which uses
a real space multiple-scattering Green's function (RSGF)
formalism to describe
the single particle spectrum and a quasi-boson model Hamiltonian 
to account for multi-electron (e.g. shake-up and shake-off)
excitations.\cite{campbell02,kas07} Although extensions are possible,
our approach is currently limited to systems with weak correlations.
Our derivation is similar in some respects
to that of Ref.~\onlinecite{fujikawa2004},
which also uses a multiple-scattering formalism and a related treatment
of inelastic losses. 
The main differences are: i) our expression does not rely on
a single site approximation
for the single particle Green's function; ii) we include
quasi-particle self-energy corrections based on a many-pole model of
the dielectric function; and iii) we approximate the
many-body losses via a convolution with an effective spectral function
that includes intrinsic and extrinsic losses and interference
effects.\cite{kas07} 
In addition, with several simplifying assumptions, we demonstrate that
the RIXS cross section can be approximated as a convolution of 
XAS and XES signals. This simplified formula is analogous to
an expression in terms of appropriate joint densities of states.
\cite{jimenez1999,glatzel2010} 
However, our result also
explicitly includes the energy dependence of the dipole matrix elements.
In order to calculate our approximation to the RIXS spectrum efficiently, we
have implemented the theory in an extension of the real space
multiple-scattering Green's function code FEFF9.\cite{rehr00,rehrpccp}
This  RSGF technique has already been used to calculate several other
core-level spectroscopies, including XAS, XES and EELS, as well
as VIS-UV spectra.\cite{rehrpccp,jorissen2010,feff84ref,prange2009} The
RSGF method has been particularly beneficial for the core-level spectroscopies
of complex systems, since it does not rely on
periodic symmetry, which is generally broken by the presence of the core hole.
Moreover, the approach is applicable over a broad range of energies.
Illustrative examples are presented
which yield reasonable agreement with available experimental RIXS.
Our theory also yields simulations for related spectra, e.g.,
 lifetime broadening suppressed x-ray absorption spectra, as observed
in high energy resolution fluorescence detection (HERFD) experiments.

The remainder of this article is ordered as follows. We begin by
introducing the theory of RIXS based on the Kramers-Heisenberg equation
and then summarize our key results. In particular 
we derive an approximate formula for the RIXS cross section in terms of a
convolution of XES and effective XAS signals.
We then present a more detailed derivation of RIXS in terms of
quasi-particle Green's functions within the RSGF formalism.
Subsequently, we present several illustrative calculations and compare with
experimental data in a number of weakly correlated systems. 
Although our treatment is in principle more general, we restrict our attention
in this work to systems for which the quasi-particle approximation is
reasonable.  Finally, we make a number of concluding remarks.
Technical details are relegated to the Appendices.

\section{Theory}
\label{sect:theory}
\subsection{RIXS in terms of XAS and XES}
Below we briefly outline the basic theory of RIXS and describe the
key expressions used in our calculations. All quantities are expressed in Hartree atomic units ($e =
\hbar = m = 1$) unless otherwise noted.
Formally the resonant inelastic x-ray scattering double differential cross
section is given by the Kramers-Heisenberg formula\cite{kramers1925}
\begin{align}
  \frac{d^2\sigma}{d\Omega d\omega} = &\frac{\omega}{\Omega} \sum_{F} \left|\frac{\sum_{M} \langle F
  | \Delta_{2}^{\dagger}|M\rangle \langle M
  | \Delta_{1}|\Psi_{0}\rangle}
  {E_{M}-\Omega-E_{0}+i\Gamma_{M}}
  \right|^{2}  \nonumber \\
  ~& \times \delta(\Omega-\omega+E_{0}-E_{F}).
  \label{eq:khrixs}
\end{align}
Here $\Omega$ and $\omega$ are the energies of the incoming and
outgoing photons; $\Delta_{1}$ and $\Delta_{2}$  are the many-body
transition operators; and $|\Psi_{0}\rangle$, $|M\rangle$ and $|F\rangle$ are
many-body electronic ground, intermediate, and final states with
corresponding energies
$E_{0}$, $E_{M}$, and $E_{F}$.
This formula for the cross section can be expressed in terms of 
effective one particle Green's functions [cf.\
Ref.~\onlinecite{fujikawa2004} and our derivation in
Sec.~(\ref{subsect:mb})] corresponding to the intermediate
and final many-body states
\begin{align}
 \label{eq:rixs_ggg}
  \frac{d^2\sigma}{d\Omega d\omega} &= -\frac{1}{\pi}\frac{\omega}{\Omega}
  |\langle b|d_{2}Q|c\rangle|^{2}
  {\rm Im} \left[\langle b|d_{1}^{\dagger}P g^{b}(\Omega+E')
  \vphantom{P}\right.\nonumber \\
  &~\times g^{c}(\Omega-\omega+E_{c})g^{b}(\Omega +
  E_{b})^{\dagger} P d_{1}|b\rangle \big].
\end{align}
Here the one-particle Green's functions $g^{b}$ and $g^{c}$ are given by
\begin{align}
    g^{b}(\omega) &= \langle \Phi^{b}_{0}|\frac{1}{\omega - h^{b}_{p} - V_{pv}
    + i\Gamma_{b}}|\Phi^{b}_{0}\rangle \nonumber \\
    &\equiv \frac{1}{\omega - h^{b}_{p} -
    \Sigma_{p}(\omega) + i\Gamma_{b}} \nonumber \\
    g^{c}(\omega) &= \frac{1}{\omega - h^{c}_{p} -
    \Sigma_{p}(\omega) + i\Gamma_{c}}\ ,
\end{align}
where $b$ and $c$ denote deep and shallow core holes,
$|\Phi^{b}_{0}\rangle$ is the ground state of the valence electrons in the
presence of core hole $b$, the projection operators $P$ and $Q$ project
onto unoccupied or
occupied states of the single particle ground state Hamiltonian, and
$E_{b}$, $E_{c}$ are the core level energies.
As shown in Appendix~\ref{sect:local}, we can rewrite this expression in terms of a
non-local transition operator $T$ 
\begin{align}
  \frac{d^2\sigma}{d\Omega d\omega} &= -\frac{1}{\pi}\frac{\omega}{\Omega}
  \frac{|\langle b|d_{2}Q|c\rangle|^{2}}{|\omega + E_{b} - E_{c} + i \Gamma_{b}|^{2}}\nonumber\\
&~~\times\, {\rm Im} \left[
\langle b|T^{\dagger}(\Omega)g^{c}(\Omega-\omega+E_{c})T(\Omega)
|b\rangle \right]  ,
\end{align}
where $T(\Omega) = \left[1 + \Delta Vg^{b\dagger}(\Omega+E_{b})\right]P d_{1}.$
We now relate this result to the x-ray emission $\mu_{e}(\omega)$ and an x-ray
absorption like signal $\bar\mu(\Omega,\Omega - \omega)$.
\begin{equation}
  \label{eq:rixsmubar}
  \frac{d^2\sigma}{d\Omega d\omega} = \frac{\omega}{\Omega}
\int d\omega_{1}~\frac{\mu_{e}(\omega_{1})
    \bar\mu(\Omega, \Omega - \omega - \omega_{1} + E_{b})}{|\omega - \omega_{1} + i\Gamma_{b}|^{2}}.
\end{equation}
where the effective absorption coefficient $\bar \mu$ is
\begin{equation}
  \label{eq:mubar}
  {\bar \mu}(\Omega,\Omega-\omega) =-\frac{1}{\pi} {\rm Im}\left[\langle b |
T^{\dagger}(\Omega)g^{c}(\Omega-\omega+E_{c})T(\Omega) | b \rangle\right] .
\end{equation}
The quantity $\bar\mu$ differs from normal x-ray absorption coefficient
since the dipole transition operator in XAS is replaced by $T(\Omega)$.
If the matrix elements of $g' \Delta V$ are much smaller than unity,
which is the case for all but localized excitations, we may take the
leading order approximation, and thus
relate the RIXS to the usual x-ray absorption coefficient $\mu(\omega)$, i.e.,
\begin{equation}
  \label{eq:rixsxasxes}
  \frac{d^2\sigma}{d\Omega d\omega} \propto \frac{\omega}{\Omega}
\int d\omega_{1}~\frac{\mu_{e}(\omega_{1})
    \mu(\Omega - \omega - \omega_{1} +
E_{b})}{|\omega - \omega_{1} - i\Gamma_{b}|^{2}}.
\end{equation}
Thus we obtain a relatively simple expression for the RIXS cross section in terms
of the x-ray absorption, x-ray emission, and a resonant denominator. 
Moreover, the terms in either expression [Eq.~(\ref{eq:rixsmubar}) or
(\ref{eq:rixsxasxes})] can be calculated within the
RSGF framework, as in the FEFF codes.\cite{rehr00,feff82ref,feff84ref,feff9ref,rehrpccp}
It should be noted that the above expressions (Eq.~\ref{eq:rixsmubar}
and \ref{eq:rixsxasxes}) are similar to those given in the pioneering
work of Tulkki and \r{A}berg,\cite{Aberg1980,tulkki1980,tulkki1982} in which a
derivation of electronic resonant
Raman spectra is given in terms of multichannel scattering states, and
applied to the K-alpha RIXS of KMnO$_{4}$.

\subsection{Multiple Scattering Theory}
We now turn our attention to the application of the multiple-scattering
RSGF formalism to Eq.~(\ref{eq:rixsmubar}).
Within this formalism the single particle Green's function can be expanded
about the absorbing atom (see Ref.~\onlinecite{rehr00})
\begin{align}
  G(\bm r, \bm r', E) =& -2k\left[\vphantom{\sum_{LL'}}\right.  \left.\sum_{LL'}|R_{L}(E)\rangle
 G_{L0L'0}(E) \langle R_{L'}(E)| \right. \nonumber \\
  & \left.  + ~~
\delta_{L,L'}|H_{L}(E)\rangle\langle R_{L}(E)|\vphantom{\sum_{LL'}}\right].\label{eq:msgf}
\end{align}
Calculations of
the RIXS cross section require both the single particle XES signal,
which is relatively simple to calculate with FEFF9, and the effective absorption
cross-section $\bar \mu$.  In order to calculate the latter
we begin by rewriting Eq.~\eqref{eq:mubar} in terms of the
one-electron density matrix $\rho^c$, 
\begin{equation}
  {\bar \mu}(\Omega,\Omega-\omega) \propto \langle b | T^{\dagger}(\Omega)\rho^{c}(\Omega-\omega+E_{c})T(\Omega) | b \rangle.
\end{equation}
Note again that $|b\rangle$ and $|c\rangle $ signify states
calculated in the presence of the deep or shallow core hole respectively.
Using the fact that $\rho = -(1/\pi)\, {\rm Im}\, [g]$ and inserting our
expression for the Green's function, we obtain
\begin{align}
  {\bar \mu}(\Omega,\Omega-\omega) &= -2k \sum_{LL'}\langle
  b|T^{\dagger}(\Omega)|R^{c}_{L}\rangle\, [\delta_{LL'} +\nonumber \\
   & ~
  \rho^{c}_{0L0L'}(\Omega-\omega + E_{c})]\, \langle R^{c}_{L'}|T(\Omega)|b
  \rangle,
\end{align}
where we have assumed that the argument $\Omega - \omega + E_{c}$ is real, and
used the result
${\rm Im}[|R \rangle\langle H|] = |R \rangle\langle R|$
for energies on the real axis. Note that the energy arguments of the
bras and kets have been omited above for the sake of brevity. We now 
  turn to the matrix elements of the transition operator
\begin{equation}
  T_{Lb}(\Omega) = \langle R^{c}_{L}|T(\Omega)|b \rangle = \langle 
  R^{c}_{L}|\left[\Delta Vg^{b}(\Omega)^{\dagger} + 1 \right]d_{1}|b \rangle,
\end{equation}
where for simplicity, we have neglected the projection operator $P$, i.e.
approximated $P$=1. Then rewriting the Green's function $g^{b}$ in spectral
representation, and again inserting the MS expression for the Green's
function in Eq.~(\ref{eq:msgf}) gives
\begin{align}
  \label{eq:tlb} 
  &T_{Lb}(\Omega) =    \langle R^{c}_{L}|d_{1}|b\rangle +\pi \int d\omega_{1}
  \frac{2 k_{1}}{\omega_{1} + i \Gamma_{b}}\times\nonumber \\
  &~~\sum_{L_{1}} \langle R^{c}_{L}|\Delta V|R^{b}_{L} \rangle
   \left[\delta_{LL_{1}} + 
   \rho^{b}_{LL_{1}}(\Omega-\omega_{1})\right]\langle R^{b}_{L_{1}}| d_{1}|b \rangle.
\end{align}
Thus in addition to the usual dipole matrix elements, we have a second term
which depends on both energies in the problem, the incoming photon
frequency $\Omega$ and the energy loss $\Omega-\omega$.


\subsection{Many-Body Effects and the Quasi-Boson Model \label{subsect:mb}}

In this subsection we discuss the application of the quasi-boson
model\cite{hedinr1999} to calculations of inelastic loss effects in RIXS. 
Assuming that the absorption occurs from a deep core level $|b\rangle$ and
employing the dipole approximation for the transition operators
in Eq.~(\ref{eq:khrixs}) gives
\begin{align}
  \Delta_{1} &= \sum_{k} \langle k|d_{1}|b \rangle c_{k}^{\dagger}b +
  \text{h.c.} \nonumber\\
  \Delta_{2} &= \sum_{k}\langle b|d_{2}|k \rangle
  b^{\dagger}c_{k} + \text{h.c.}.
\end{align}
If we neglect exchange terms between the particle and hole, or at
least assume that they are dealt 
with via an effective single particle potential, we can write the
many-body ground state as
\begin{equation}
  |\Psi_{0}\rangle = |\Phi_{0}\rangle|b\rangle|k_{2}\rangle,
\end{equation}
where $k_{2}$ is associated with a specific term in the sum over
states in $\Delta_{2}$, $|b\rangle$ is the deep core state excited by
the absorption event, and $|\Phi_{0}\rangle$ is an $N-2$ electron
wave-function. Note that this approximation is only justified if $k_{2}$ denotes a core electron or a high energy photo-electron, although we will use the approximation for valence electrons as well. This gives
\begin{align}
  \Delta^{k_{1}}_{1}&|\Psi_{0}\rangle =
  M_{1}^{k_{1} b}|\Phi_{0}\rangle|k_{2}\rangle|k_{1}\rangle\theta(E_{k_{1}}-E_{\rm F}) \nonumber \\
  \Delta^{k_{2}}_{2}&|\Phi_{0}\rangle|k_{2}\rangle|k_{1}\rangle =
   M_{2}^{b k_{2}}|\Phi_{0}\rangle|b\rangle|k_{1}\rangle \theta(E_{\rm F} - E_{k_{2}}),
\end{align}
where $M_{i}^{k b} = \langle k | d_{i} | b \rangle$ and
\begin{equation}
  H|\Psi_{0}\rangle = E_{0}|\Psi_{0}\rangle = (\epsilon_{b} +
  \epsilon_{k_{2}} + E_{0}^{0})|\Psi_{0}\rangle.
\end{equation}
Note that $E_{\rm F}$ is now the Fermi energy. Then the RIXS cross section becomes
\begin{align}
  \frac{d^2\sigma}{d\Omega d\omega} &= -\frac{1}{\pi}\frac{\omega}{\Omega}
       {\rm Im}\left[\sum^{\rm unocc}_{k_{1} k_{2}}\sum^{\rm occ}_{k_{3} k_{4}} M_{2}^{b k_{3}} (M_{2}^{k_{4} b}M_{1}^{b k_{1}})^{*}M_{1}^{k_{2} b}
       \right. \nonumber \\
        &\langle k_{1}| \langle k_{3}| \langle \Phi_{0}|
K_{k_{3}k_{4}}(\xi_{1},\xi_{2})
\left.|\Phi_{0}\rangle|k_{4}\rangle|k_{2}\rangle
  \vphantom{\sum^{\rm unocc}_{k_{1} k_{2}}}\right],
\end{align}
where $\xi_1 = \Omega+E_0$, $\xi_2 = \Omega+ E_0 - \omega$, and
$K$ is given by
\begin{equation}
  K_{k k'}(\omega,\omega') =  G(\omega) c^{\dagger}_{k} b G(\omega') b^{\dagger} c_{k'} 
  G(\omega)^{\dagger},
\end{equation}
and $G(E) = 1/(E-H+i\delta)$ is the many-body Green's function.
We now introduce a quasi-boson approximation to the Hamiltonian following
the treatment of Ref.~\onlinecite{campbell02}. In this approach
the excitations of the many-body valence electronic state are represented
as bosons while the photo-electron and hole are treated via an
effective single-particle theory
\begin{equation}
  H = H_{0}^{N-2} + h_{p} + h_{h} + V_{hv} + V_{pv} + V_{ph} ,
\end{equation}
where $h_{h}$ and $h_{p}$ are the one-particle Hamiltonians for the
hole and particle respectively, $V_{hv}/V_{pv}$ describes the interaction of the
hole/particle with the valence electrons, 
\begin{align}
  h_{p} &= \sum_{k}\epsilon_{k}c_{k}^{\dagger}c_{k};\quad h_{h} =
  -\sum_{k}\epsilon_{k}c_{k}c_{k}^{\dagger}, \\
  V_{pv} &= \sum_{n,k_{1}k_{2}}\left[V^{n}_{k_{1}k_{2}}\adag{a}{n}
    + (V^{n}_{k_{1}k_{2}})^{*}a_{n}\right]\adag{c}{k_{1}}c_{k_{2}}, \\
  V_{hv} &= \sum_{n,k_{1}k_{2}}\left[V^{n}_{k_{1}k_{2}}\adag{a}{n}
    + (V^{n}_{k_{1}k_{2}})^{*}a_{n}\right]c_{k_{1}}\adag{c}{k_{2}},
\end{align}
and $V_{ph}$ describes the interaction between the photo-electron and
hole. This last interaction term should in principle
be treated via the Bethe-Salpeter
equation; however, here we will approximate it using either a self-consistent
final state rule approximation for deep core holes (i.e., with
the screened core-hole potential of the deep core-hole),
or by neglecting it altogether, as in the initial state rule
(independent particle approximation)
for valence holes. Experience with such models in the FEFF
code shows that these approximations are reasonable.

We now define 
 $|\Phi^{b}_{0}\rangle$ as the ground state of the $N-2$ electron
system in the presence of the deep core hole $|b\rangle$, and
$|\Phi^{c}_{0}\rangle$ as the ground state of the $N-2$ electron
system in the presence of the second core hole $|c\rangle$ so that
\begin{align}
  H^{b}|\Phi^{b}_{0}\rangle &= E^{b}_{0}|\Phi^{b}_{0}\rangle; ~~H^{b} = H_{0}^{N-2} +
  V^{b}_{hv} \nonumber \\
  {H^{c}}|{\Phi^{c}_{0}}\rangle &= {E^{c}_{0}}|
  \Phi^{c}_{0}\rangle; ~~H^{c} = H_{0}^{N-2} +
  V^{c}_{hv}i\ ,
\end{align}
with core level energies 
\begin{align} 
E_{b} &= \epsilon_{b} - E_{0}^{0} + E^{b}_{0},\nonumber \\
E_{c} &=\epsilon_{c} - E_{0}^{0} + E^{c}_{0}.
\end{align}
The
transition matrix elements corresponding to emission ($d_{2}$) may be
pulled outside the imaginary part, and serve as an amplitude factor, i.e.,
\begin{align}
  \frac{d^2\sigma}{d\Omega d\omega} &= -\frac{1}{\pi}\frac{\omega}{\Omega}
       \sum_{c} |\langle b|d_{2}Q|c \rangle|^{2}\nonumber\\
       &\times \, {\rm Im}\left[\sum^{\rm unocc}_{k_{1} k_{2}}
       \langle b |d_{1}^{\dagger} P
      F(E_1,E_2) P d_{1}|b\rangle
       \right].
\end{align}
Here $E_1 = \Omega+E_b$, $E_2=\Omega+E_c-\omega$,
$P$ is a projector onto unoccupied states of the ground state
Hamiltonian, $Q$ is a projector onto occupied states of the intermediate
state Hamiltonian,  
\begin{equation}
  F(E_{1},E_{2}) =  G^{b}(E_{1}) G^{c}(E_{2}) G^{b\dagger}(E_{1});
\end{equation}
and finally, the Green's functions are calculated in the presence of the deep
($b$) or shallow ($c$) core hole, i.e.,
\begin{align}
  G^{b}(\omega) &= \frac{1}{\omega - (H_{b}-E^{b}_{0}) - h^{b}_{p}- V_{pv} +
  i\Gamma_{b}} \nonumber \\
  G^{c}(\omega) &= \frac{1}{\omega - (H_{c}-E^{c}_{0})
  - h^{c}_{p} - V_{pv} + i \Gamma_{c}},
\end{align}

Next we derive  an expression for the effects of multi-electron excitations
in terms of an effective spectral function. 
Within the quasi-boson approximation, the following relationships
between the eigenstates of $H_{0}$, $H^{b}_{0}$,
and $H^{c}_{0}$
hold\cite{kas07,campbell02}
\begin{align}
  |\Phi_{0}\rangle &= e^{-S_{b}}|\Phi^{b}_{0}\rangle; 
  ~~~S_{b} = \frac{a_{b}}{2} - \sum_{n}\vbbnoon{n} \adag{a}{bn}; \nonumber \\
  |\Phi^{b}_{0}\rangle &= e^{-\Delta S}|\Phi^{c}_{0}\rangle; 
  ~~{\Delta S} = \frac{\Delta a}{2} - \sum_{n}\deltavoon{n} \adag{a}{cn};
   \nonumber \\
   {\Delta a} &= \sum_{n}\left(\deltavoon{n}\right)^{2},\ 
    a_{b} = \sum_{n}\left(\vbbnoon{n}\right)^{2} .
\end{align}
Here $\Delta V^{n} = V^{n}_{cc} - V^{n}_{bb}$ is the difference between
the intermediate and final state core hole potentials. If we assume only single boson excitations, we can also write
\begin{equation}
  |\Phi^{b}_{n}\rangle  = \left[\adag{a}{cn} -
    \deltavoon{n}\right]e^{-\Delta S}|\Phi^{c}_{0}\rangle,
\end{equation}
which will give us the correct expression to second order in the couplings when used in our formula for the RIXS signal.
Ignoring the off-diagonal terms in $V_{pv}$ 
we obtain
\begin{align}
  \frac{d^2\sigma}{d\Omega d\omega} &= -\frac{1}{\pi}\frac{\omega}{\Omega}
  {\rm Im} \left\{ \sum_{n_{1}n_{2}}
  \langle b|d_{1}^{\dagger}P \langle
  \Phi^{b}_{0}|e^{-{S_{b}}^{\dagger}}G^{b}(E_1) |\Phi^{b}_{n_{1}} \rangle \right. \nonumber \\
  &~
  \langle b|d_{2}Q \langle \Phi^{c}_{0}| e^{-\Delta S^{\dagger}} \left[{a}_{c n_{1}} - \left(\deltavoon{n_{1}}\right)^{*}\right] \nonumber \\
  & ~\left. G^{c}(E_2)
  \left[a^{\dagger}_{c n_{2}} -
  \deltavoon{n_{2}}\right]e^{-\Delta S}|\Phi^{c}_{0}\rangle Q
  d^{\dagger}_{2}|b \rangle  \right. \nonumber\\ 
  & ~\left.\langle \Phi^{b}_{n_{2}}| \left[G^{b}(\Omega +
    E_{c})\right]^{\dagger}e^{-{S_{b}}}|\Phi^{b}_{0}\rangle P d_{1}|b\rangle 
  \vphantom{\sum^{\rm unocc}_{k_{1} k_{2}}}\right\}.
  \label{eq:mbrixs}
\end{align}
Note that in the case of valence
emission (valence hole), we are assuming that the core hole potential
is negligible, hence $E_{c} =
\epsilon_{c}$, and $\Gamma_{c} = 0$.
Expanding to second order in the amplitudes to create and annihilate
bosons, and neglecting off resonant terms, gives the total cross
section in terms of a convolution with an effective spectral function $A_{\rm eff}$.
\begin{align}
  \frac{d^2\sigma}{d\Omega d\omega} =&
\int\,d\omega_{1}d\omega_{2}\,
A_{\rm eff}(\Omega,\Omega-\omega,\omega_{1},\omega_{2})
 \nonumber \\
 &\times \left. \left[\frac{d^2\sigma}{d\Omega d\omega}\right]_{\rm sp}
\right|_{\Omega = \Omega - \omega_{1},\omega =
\omega - \omega_{1}+\omega_{2}} ,
\label{eq:rixs_conv}
\end{align}
where $\left[{d^2\sigma}/{d\Omega d\omega}\right]_{\rm sp}$ is the
single particle cross section as given in Eq.~(\ref{eq:rixsmubar}),
and the spectral function is given by
\begin{align}
&A_{\rm eff}(E_{1},E_{2},\omega_{1},\omega_{2}) = e^{-a_{c}}
  \left\{\vphantom{\sum_{n}\left[\beta_{cn}(E_{2})\alpha_{cn}(E_{2})\delta(\omega_{1})\delta(\omega_{2} - \omega_{n})
      \right.}
    \delta(\omega_{1})\delta(\omega_{2}) \right. \nonumber \\
    &~~ +\, \sum_{n}\left[\beta_{cn}(E_{2})\alpha_{cn}(E_{2})\delta(\omega_{1})\delta(\omega_{2} - \omega_{n})
    \right.\nonumber\\
    & + \left.\left.|\beta_{bn}(E_{1})|^{2}\delta(\omega_{1} -
    \omega_{n})\delta(\omega_{2} -
  \omega_{n})\vphantom{f_{cn}}\right]\vphantom{\sum_{n}}\right\}.
    \label{eq:mbspectfn}
\end{align}
Here $\alpha_{n}$, $\beta_{n}$ are the amplitudes to create or
annihilate a bosonic excitation, respectively, and include extrinsic
as well as intrinsic amplitudes (see Appendix~\ref{sect:sf}). It should be noted that our current
formalism for the spectral 
function is not suited for highly correlated materials, although an
extension of the quasi-boson model Hamiltonian is possible, as
suggested in Ref.~\onlinecite{hedinr1999} and \onlinecite{hedin1999}.

The application of the spectral function to RIXS is similar to 
that of Ref.~\onlinecite{campbell02} and
\onlinecite{kas07}, where a convolution was applied to XAS.
In this  paper, however, we will restrict our calculations to the
quasiparticle approximation, i.e., with the spectral function replaced
by a $\delta$-function,
\begin{equation}
A_{\rm eff}(E_{1},E_{2},\omega_{1},\omega_{2}) =
\delta(\omega_{1})\delta(\omega_{2} - \omega_{n}).
\label{eq:qpspectfn}
\end{equation}
The use of this approximation is expected to cause the calculated spectral
line-shapes to be more symmetric than experimental results, since the main
quasiparticle peak is modeled in the above as a Lorentzian, while  
in general multi-electron excitations lead to asymmetric peaks so that
Eq.~(\ref{eq:mbspectfn}) has a Fano type main lineshape. Thus
satellite peaks due to multi-electron excitations are also neglected.
\section{Experiment}
The experiments described here were performed at beamline ID26 of the European
Synchrotron Radiation Facility (ESRF). The incident energy was
selected by means of a pair of cryogenically cooled Si crystals in
($311$) reflection with an energy bandwidth of $0.2$ eV ($0.3$ eV) at
$4.9$ keV ($6.5$ keV). The incident flux on the sample was
$1\times10^{13}$ photons/second using the fundamental peak of the
undulator radiation. The beam size on the sample was $0.2$ mm vertical
by $1.0$ mm horizontal. Higher harmonics were suppressed by three Si
mirrors operating in total reflection. The resonantly scattered x-rays
were analyzed using the ($331$) and ($400$) reflection of spherically
bent Ge single crystal wafers for the Ti $K_{\beta}$ and $K_{\alpha}$
emission, respectively. The Ge ($333$) reflection was used for Mn
$K_{\alpha}$. Sample, analyzer crystals and an avalanche photo diode
were arranged in a vertical Rowland geometry ($R=1$ m) at $90 \pm 3$
deg scattering angle. The combined instrumental energy bandwidth
was $0.8 - 1.0$ eV. All samples were purchased from Aldrich
and used as is. Self-absorption effects distort the spectral shape and
let the K absorption pre-edge region appear stronger relative to the
edge jump. These effects are negligible in the K absorption pre-edge
region and the samples were not diluted for the measurements. 

\begin{figure}
  \includegraphics[height=\columnwidth, angle = -90]{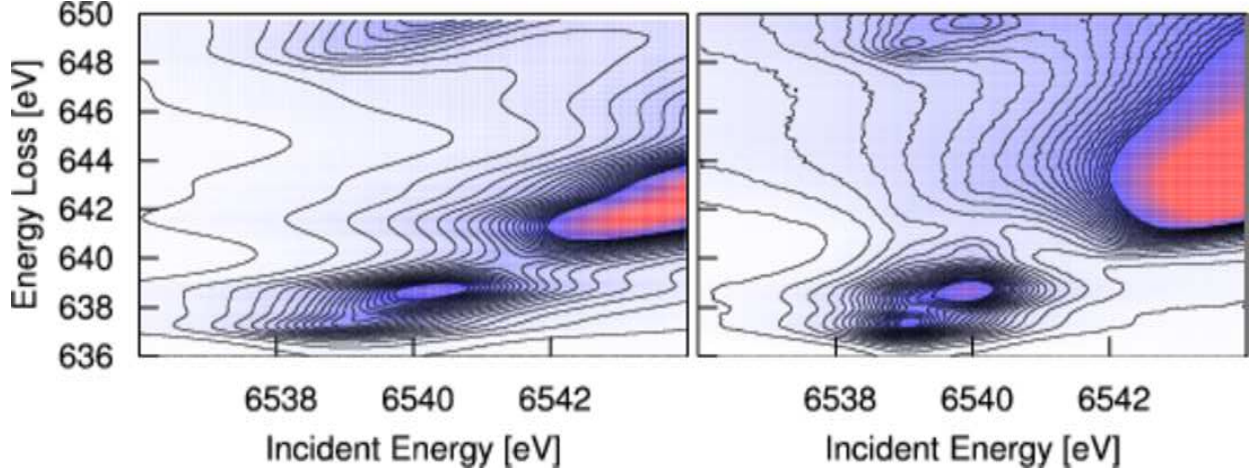}%
  \caption{(color online) Calculated (left) Mn ${\rm K}_\alpha$
  RIXS of MnO based on Eq.~(\ref{eq:rixs_conv})
  compared to experiment (right).\cite{glatzel04}
 (right).\label{fig:mno}}
\end{figure}
\section{Results and Discussion}
\subsection{RIXS}

\begin{figure}
  \includegraphics[height=\columnwidth, angle = -90]{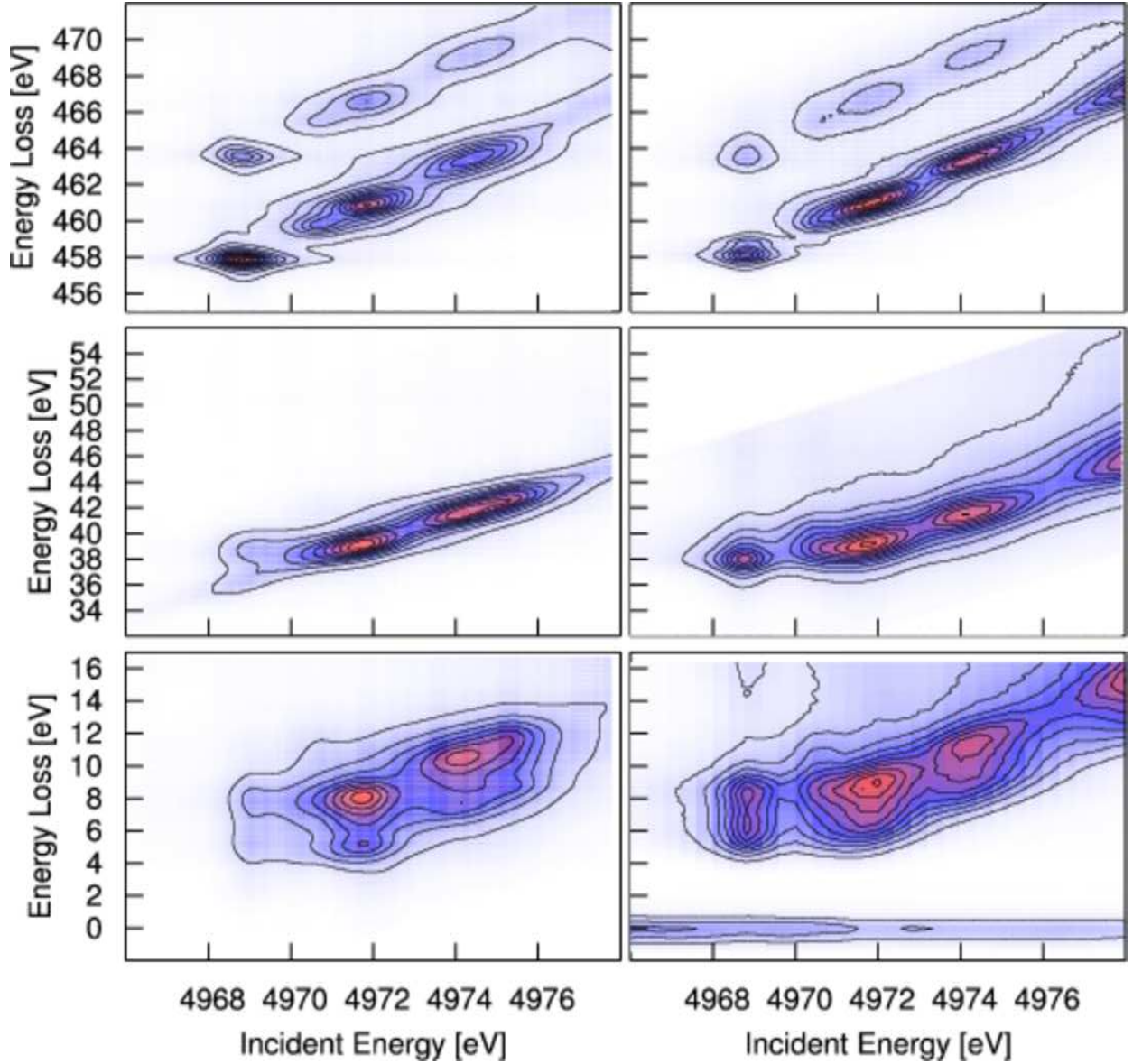}%
  \caption{(color online) Calculated RIXS (left) of ${\rm TiO}_{2}$
  based on Eq.~(\ref{eq:rixs_conv}) compared to experiment (right).
  for (top to bottom) Ti ${\rm K}_\alpha$,
  ${\rm K}_\beta$, and $K_{\rm valence}$ RIXS.} 
\end{figure}


Calculations of RIXS were carried out for several materials based on
the present theoretical approach using an extension of the RSGF FEFF9
code applied to Eq.~(\ref{eq:rixs_conv}).
All results were calculated using self-consistent potentials and a
full multiple scattering (FMS) treatment of the Green's functions for
suitably large clusters centered at the core absorption site. The core-hole
screening was calculated using the random phase approximation
(RPA),\cite{ankudinovBSE} and non-spherical parts of the core-hole
potential were neglected. In order to simplify the calculations 
the Green's functions are restricted to include only the site and
angular momentum diagonal elements. We find that for the cases
presented here, the angular momentum diagonal
approximation is reasonable for all but the lowest energy peaks since
the overlap with $\Delta V$ is then small, and our approximation then
becomes equivalent to Eq.~(\ref{eq:rixsxasxes}) in terms of XES and XAS where the angular momentum diagonal elements dominate due to dipole selection rules. 
To obtain better agreement with
the the experimental threshold energy, we allowed a small shift of the
calculated Fermi energy, which is typically too high by about 1~eV in
the
self-consistent
FEFF9.0 calculation. In addition, overall energy shifts were added
in each axis in order to align the calculation with
the energy scale in the experiment. In the case of ${\rm K}_\alpha$
RIXS, atomic values were used for the splitting between the ${\rm P}_{1/2}$ and ${\rm P}_{3/2}$
emission energies.\cite{bearden67} In addition the amplitudes were taken
(from simple counting arguments) to have a ratio $A_{3/2}/A_{1/2} =
2$, although this ratio is not generally accurate, since the particle-hole
interaction mixes the hole states.
Fig.~\ref{fig:mno} presents a comparison of our calculated (right)
Mn ${\rm K}_\alpha$ RIXS of MnO and experimental data (left).
The overall agreement is qualitatively satisfactory: all main features
of the experiment are reproduced including both the dipole
as well as quadrupole pre-edge peaks. The main edge is also at
about the correct energy, although the asymmetry caused by multi-electron
excitations is absent in our calculation which is restricted to the
quasiparticle level where the spectral function is given by
Eq.~(\ref{eq:qpspectfn}). We expect that going beyond this
quasi-particle approximation for the
spectral function as in Eq.~\ref{eq:mbspectfn} would
capture some of the asymmetry since the Lorentzian spectral shape
currently used for the quasiparticle peak in these calculations would
be replaced by a Fano type lineshape.
In addition, new features could arise due to
satellite peaks in the spectral function. The main diagonal
structure in our 
calculation appears to be sharper that that of the experiment; this
is possibly due to self-absorption effects in the experimental data.
Note that the pre-edge peak for this case is basically on the diagonal;
i.e., the emission energy is the same for the pre-edge peak as for the main
edge.

Core-hole effects are important for a variety of excited state
spectoscopies, including EELS and XAS as well as RIXS.\cite{glatzel2009,cabaret1999} RIXS spectra in
particular however, can give us insight into these effects since the
intermediate and final states have different
core-holes.\cite{glatzel2009a} 
In order to illustrate the effect of different final state core-holes,
we calculated the RIXS for Ti ${\rm K}_\alpha$, ${\rm K}_\beta$, and
${\rm K}_{\rm valence}$
RIXS of TiO$_{2}$ Anatase. In order to obtain reasonable results for
the quadrupole peak in the K-edge absorption, we increased the strength of
the core-hole potential by using $95 \%$ screened core-hole and
$5 \%$ bare core-hole,
and kept the same ratio for all core-hole calculations. Note that our
calculation again reproduces all peaks, although the intensity of the
calculated quadrupole peaks is weak compared to that observed
in the experiment.  There is also a noticeable effect
on the spectrum due to changes in final state core-hole. For the
${\rm K}_\alpha$ RIXS, the intermediate ($1{\rm s}$) and final ($2{\rm p}$)
core-holes are both quite localized
and the difference $\Delta V$ is small, thus peaks
should occur roughly on the diagonal, as seen by Eq.~(\ref{eq:tlb}).
For the ${\rm K}-\beta$ spectrum the final state has a $3{\rm p}$ core hole,
and the core-hole potential has a
vastly different shape than the $1s$ core-hole potential. This
causes $\Delta V$ to be large, and we expect off diagonal peaks to be
present. This is indeed the case, although only the quadrupole peaks
are off diagonal. This is due to the fact that the dipole pre-edge
peaks are caused by $p-d$ hybridization between the absorbing atom and
neighboring Ti atoms, and thus are relatively unaffected by the
core-hole potential. The quadrupole peaks, however, are due to a direct
transition to the Ti $d-$states which are localized around the
absorbing atom and are effected by the core-hole potential to a
greater extent than the hybridized $p$-states. The effect is also
present in the valence spectrum. In addition, the valence spectrum has multiple
peaks due to the fact that the emission is from a broad valence band
which is split due to solid state effects. The qualitative structure
of the valence band is also correct in the calculation, which reproduces
the double peak structure with the correct splitting. The intensities
are also qualitatively correct with the lower energy-transfer peaks
being less intense than the higher energy-transfer peaks. The gap is
too small in our calculation, however, this could be accounted for via a GW
gap correction. Note that we have not included the elastic scattering
contribution in our calculation of the valence RIXS, which
could effect the overall asymmetry of the signal.
Finally, for all three spectra, the main edge occurs at a larger energy
in the calculated results than in the experiment.
This could be due to strong correlation effects, which would be
expected to shift the Ti $d$-states closer to the
$p$-states. Another possible explanation which is important in the
case of Ti K pre-edge XAS of Rutile ${\rm TiO}_2$ is the failure of
the spherical muffin-tin approximation.\cite{cabaret1999}

\subsection{Lifetime Broadening Suppressed XAS}
\begin{figure}
  \includegraphics[angle=-90,width=\columnwidth]{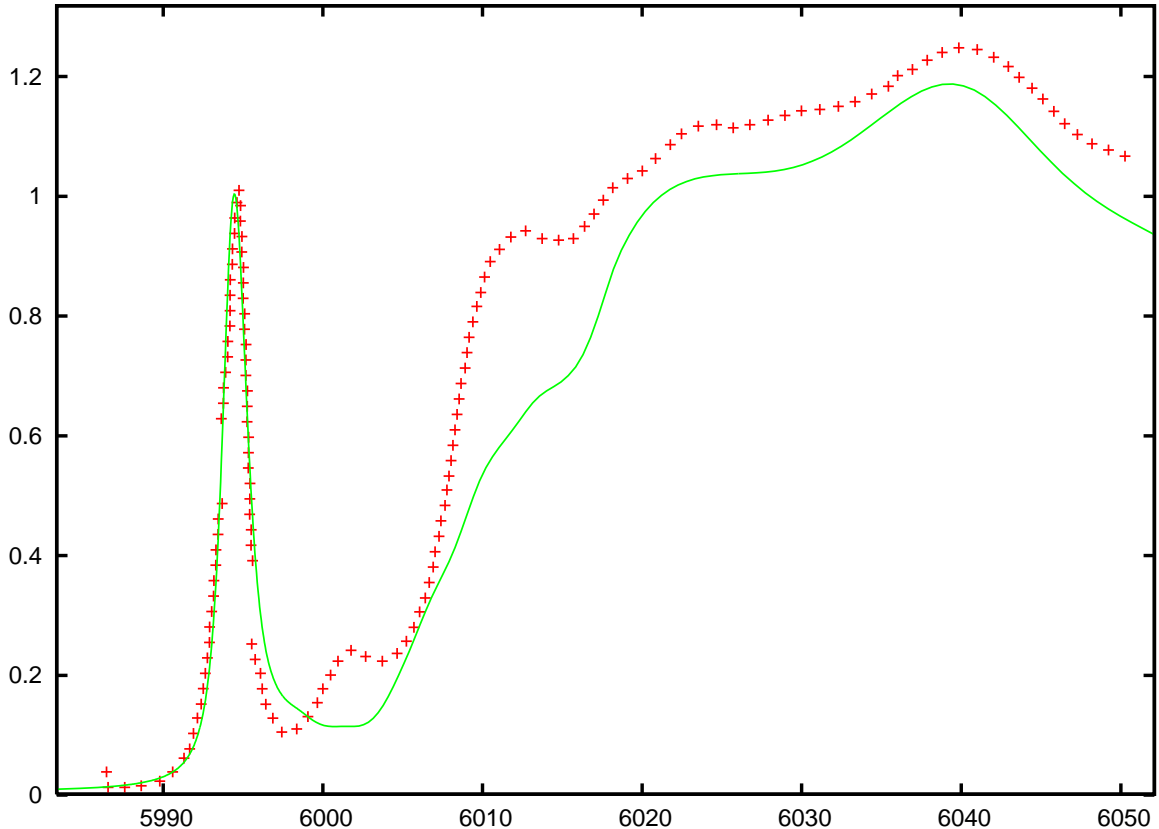}%
  \caption{(color online) Experimental (crosses) Cr K-edge HERFD XAS\cite{tromp_pc} of ${\rm K}_{2}{\rm
  CrO}_{4}$  compared
  to our calculated results (solid).\label{fig:k2cro4herfd}}
\end{figure}

In addition to the RIXS planes, there are also several methods for
obtaining lifetime broadening suppressed (LBS) XAS. In high energy
resolution fluorescence detected (HERFD) XAS,\cite{hamalainen91} an
approximate
absorption spectrum is found by partial fluorescence yield using a
detector with resolution higher than the natural width due to
core-hole lifetime effects. This
corresponds to viewing the spectra of constant emission energy in the
RIXS plane. Another method of obtaining LBS XAS is to set the incident
energy at a point well below the edge while scanning the emission
energy.\cite{hayashi03} Under certain assumptions, the spectrum obtained in this way is
approximately proportional to the XAS signal multiplied by a
Lorentzian with the width of the intermediate state core-hole.
In Fig.~\ref{fig:k2cro4herfd} we show a comparison of experimental
Cr-K edge HERFD XAS of ${\rm K}_{2}{\rm
  CrO}_{4}$ with our calculated results. We find reasonable
qualitative agreement, with the exception of the peak just above
$6000$ eV, which is not seen in the calculation. We note however, that
the size of this peak is sensitive to distortions. In addition, the
amplitude of the main peak
after the rising edge is too small. This could be due to the
approximate treatment of the core-hole interaction or corrections to
the spherical muffin-tin potentials used in FEFF9.
\begin{figure}
  \includegraphics[angle=-90,width=\columnwidth]{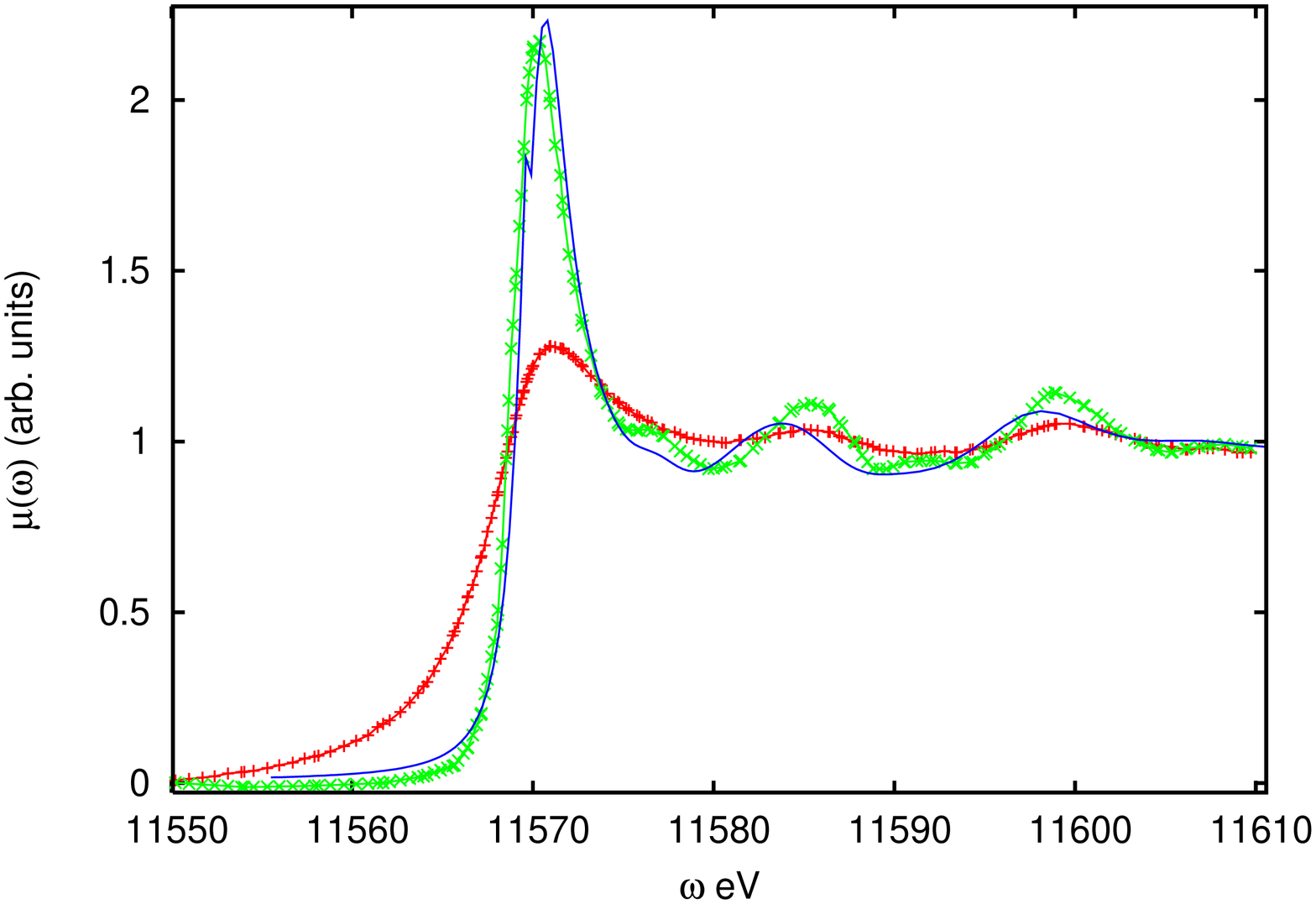}%
  \caption{(color online) Pt L$_3$ edge normal XANES (crosses) compared with HERFD
  XAS (x's) and our calculated result (solid).\label{fig:ptherfd}}
\end{figure}
Fig.~\ref{fig:ptherfd} shows a comparison of bulk metallic Pt L$_3$ edge
XANES compared to HERFD XAS and our calculated results. Here we see
that all of the features are well reproduced although the broadening
is too large at high energies, and the higher energy peaks are also
red shifted toward the edge in comparison to the experimental result.

\section{Conclusions}
We have presented a theory of resonant inelastic x-ray scattering (RIXS)
which is amenable to practical calculations as an extension of current
x-ray-absorption and -emission codes.  Starting from the
Kramers-Heisenberg equation, we derive an expression for the RIXS
cross-section which can  be calculated using the real-space Green's function
approach in the FEFF9 code.
Inelastic losses and quasi-particle effects are included in terms of
an effective spectral-function that is obtained from 
a quasi-boson model Hamiltonian.  These many-body effects are 
incorporated into a single-particle approximation via a convolution with
an effective spectral function. Quasi-particle self-energy effects are
included based on a many-pole model of the dielectric function. 
Approximation of the many-body states
as a product of an N-2 electron state with either two core electronic
states (i.e. the ground state), or a core and photo-electron state
(intermediate and final states) gives the cross section in terms of
effective single particle Green's functions. The further
approximation that the intermediate and final photo-electron states
are orthogonal with identical energies (valid at high photo-electron energies)
gives the signal in terms of a convolution of the XAS and XES spectra. 
In addition, we have derived a formulation of the core-core RIXS spectrum
that, due to the localized nature of $\Delta V$, depends
primarily on the Green's functions evaluated close to the absorbing
atom. The extent of this localization is yet to be thoroughly investigated,
although the degree of agreement between our calculations and
experimental data suggests that the on-site approximation is valid for
the systems shown here. In addition, the on-site approximation provides
good qualitative agreement for the core-valence RIXS, although the
approximation is less justifiable. The theory is implemented in an efficient program which is 
 an extension of the FEFF9 real-space multiple-scattering code, which
calculates the RIXS spectrum for core-core as well as core-valence RIXS.
Several illustrative calculations were presented within the quasi-particle
approximation where the spectral function is replaced by a $\delta$-function,
which appears to be a reasonable approximation for these cases.
Calculated results for MnO and for Anatase TiO$_2$  based on this
quasi-particle approximation are found to agree qualitatively with
experimental spectra: the results reproduce both pre-edge and main edge
features, the behavior of pre-edge features with varying core-hole
interaction strength, and peak structure due to solid state effects in
valence RIXS. Further investigations including treatments beyond
the quasi-particle approximation of Eq.~(\ref{eq:qpspectfn}) will
be reserved for the future.
It should be noted that our current formalism for the spectral
function is not suited for highly correlated materials, although an
extension of the quasi-boson model Hamiltonian is possible as
suggested in Ref.~\onlinecite{hedinr1999} and \onlinecite{hedin1999}.

\acknowledgments
We thank T.~Ahmed, A. Bansil, R. Markiewicz, E. Shirley, and M.~Tromp, 
for useful discussions.
This work was supported by DOE BES Grant DE-FG03-97ER45623
and was facilitated by the DOE Computational Materials and Chemical
Sciences Network (JJK and JJR). JAS gratefully acknowledges the
financial support from 
Eemil Aaltonen foundation and Magnus Ehrnrooth foundation. The ESRF is
acknowledged for providing beamtime and technical support (PG). 


\appendix
\section{Local behavior of RIXS: Derivation of $T(\Omega)$}
\label{sect:local}
The RIXS cross section is given in terms of a product of three Green's
functions, i.e., ${\rm
Im}[g^{b}(E_{1})g^{c}(E2)g^{b}(E1)^{\dagger}]$. However, this
expression is not very useful, since the spacial arguments of the Green's
function must be integrated over all space to obtain the
resonance $1/(E1-E2)$. In order to see this, imagine that we ignore
the core hole potentials in both Green's functions. In this case, we
may write the Green's functions in spectal representation, and noting
that the wave functions are now orthonormal, we may rewrite the above
expression as $|{\rm Im}[g(E_{2})/(E1-E2+i\Gamma)|^{2}$, which is what
we expect for energies well above threshold, where the effect of the
corehole potential is negligable. While this limit is easy to show
analytically, doing so numerically within the real-space MS Greeen's
function formalism proves quite difficult.
Below, we derive an alternative expression for the
RIXS cross section which takes advantage of the localization of
core-hole potential. Let us first define the Hamiltonian operators
corresponding to the deep (b) and shallow (c) core holes.
\begin{align}
h_{c} &= h_{0} + V_{c} \\
h_{b} &= h_{0} + V_{b} = h_{c} + \Delta V \\
\Delta V &= V_{b} - V_{c}.
\end{align}
We can use these definitions to rewrite the Green's funtions as follows
\begin{align}
g^{b}[E_{1} - h_{b}] &= {\bf 1} ~\Rightarrow~ g^{b} =  \frac{{\bf 1} + g^{b}[h_{c} + \Delta V]}{E_{1}} \\
g^{c} h_{c} &= - {\bf 1} + E_{2} g^{c},
\end{align}
where we have left the energy arguments off of the Green's functions
for the sake of brevity. 
Using these relations gives
\begin{align}
g^{b} g^{c} &= \frac{g^{c} - g^{b} + g^{b} \Delta V
g^{c}}{E_{1}-E_{2}} = \frac{D^{\dagger}g^{c}-g^{b}}{E_{1}-E_{2}}, \\ 
g^{c}g^{b \dagger} &= \frac{g^{c} - g^{b\dagger} + g^{c} \Delta V
g^{b\dagger}}{E_{1}^{*}-E_{2}} = \frac{g^{c}D - g^{b \dagger}}{E^{*}_{1}-E_{2}},
\end{align}
where $D = 1+\Delta V g^{b \dagger}.$
Applying the above relations to
$g^{b}(E_{1})g^{c}(E2)g^{b}(E1)^{\dagger}$ gives

\begin{align}
&g^{b} g^{c} g^{b\dagger} = \frac{1}{2}\left[(g^{b} g^{c} )
g^{b\dagger} + g^{b}(g^{c}g^{b\dagger})\right] \nonumber \\
& = \frac{1}{2}\left[\frac{(D^{\dagger}g^{c}-g^{b})g^{b
\dagger}}{E_{1}-E_{2}} + \frac{g^{b}(g^{c} D - b^{b
\dagger})}{E^{*}_{1}-E_{2}}\right] \nonumber \\
& = \frac{D^{\dagger} g^{c} D}{|E_{1} - E_{2}|^{2}} -
\frac{1}{2}\left[\frac{D^{\dagger}g^{b \dagger}}{|E_{1} - E_{2}|^{2}} +
\frac{g^{b}g^{b \dagger}}{E_{1} - E_{2}} + {\rm h.c.}\right]
\end{align}
Noting that the second term above is real, we have
\begin{equation}
{\rm Im}\left[g^{b}g^{c}g^{b\dagger}\right] = \frac{D^{\dagger}{\rm Im}[g^{c}]D}{|E_{1} - E_{2}|^{2}}.
\end{equation}
Finally, we see that the transition matrix element $T$ defined in
Section~(\ref{sect:theory}) can be related to $D$, i.e.,
\begin{equation}
T = D P d
\end{equation}

\section{Effective Spectral function \label{sect:sfderivation}}
\label{sect:sf}
As shown in Sec.\ (\ref{sect:theory}), the RIXS cross section is given in terms of the ground state expectation value of a product of three many-body Green's functions. Here we will derive an expression based on quasi-particle Green's functions and a many-body spectral function.
\begin{align}
  \label{eq:mbggg}
  &\langle \Phi_{0}|G^{b}(E_{1})G^{c}(E_{2})G^{b\dagger}(E_{1})|\Phi_{0}\rangle  \nonumber \\
& = \sum_{n_{1}n_{2}}
  \langle\Phi^{b}_{0}|e^{-S^{b\dagger}}G^{b}(E_{1})|\Phi^{b}_{n_{1}}\rangle
\langle\Phi^{b}_{n_{1}}|G^{c}(E_{2})|\Phi^{b}_{n_{2}}\rangle \nonumber\\
&\times\hphantom{\sum_{n1 n2}}\langle\Phi^{b}_{n_{2}}|G^{b\dagger}(E_{1})e^{-S^{b}}|\Phi^{b}_{0}\rangle
\end{align}

Note that if we are expanding to second order in the boson couplings, only $|\Phi_{n}\rangle$ containing single boson excitations, $|\Phi_{n}\rangle = a_{n}^{\dagger}|\Phi_{0}\rangle$ contribute.
Now,
\begin{equation}
  |\Phi^{b}_{n}\rangle 
= Z_{n}e^{-\Delta S}|\Phi^{c}_{0}\rangle 
\end{equation}
where $Z_{n} = [a_{n}^{c\dagger} + \Delta V^{n}/\omega_{n}]$
and $\Delta V^{n} = V^{cn} - V^{bn}$. Thus Eq.~(\ref{eq:mbggg}) becomes
\begin{align}
  \langle &\Phi_{0}|G^{b}(E_{1})G^{c}(E_{2})G^{b\dagger}(E_{1})|\Phi_{0}\rangle  \nonumber \\ 
  &=\sum_{n_{1}n_{2}}
  \langle\Phi^{b}_{0}|e^{-S^{b\dagger}}G^{b}(E_{1})|\Phi^{b}_{n_{1}}\rangle \nonumber \\
&\times \langle\Phi^{c}_{0}|e^{-\Delta S^{\dagger}}Z_{n} 
  G^{c}(E_{2})Z_{n}^{\dagger} 
  e^{-\Delta
  S}|\Phi^{c}_{0}\rangle \nonumber \\
  &\times\langle\Phi^{b}_{n_{2}}|G^{b\dagger}(E_{1})e^{-S^{b}}
  |\Phi^{b}_{0}\rangle. 
\end{align}
We now define the amplitudes to create a single boson as
\begin{align}
  \alpha^{b}_{n} &= V^{n}G^{b}(E_{1}) - \vbbnoon{n}; &\alpha^{c}_{n} &= V^{n}G^{c}(E_{2}) - \deltavoon{n},
\end{align}
and the amplitudes to annihilate a boson as
\begin{align}
  \beta^{b}_{n} &= G^{b}(E_{1})(V^{n})^{*} - \vbbnoon{n}; &\beta^{c}_{n} &=
G^{c}(E_{2})(V^{n})^{*} - \deltavoon{n}.
\end{align}
Then combining Eq.\ (B3)-(B5) gives six terms quadratic in these amplitudes
plus the zeroth order term.
If we neglect the off-resonant terms, i.e. those terms which contain
a $G^{b}$, $G^{b\dagger}$ with different energy arguments, this leaves
only two terms plus the zeroth order term, 
\begin{align}
\langle \Phi_{0}&|G^{b}(E_{1})
    G^{c}(E_{2})G^{b\dagger}(E_{1})|\Phi_{0}\rangle  \nonumber \\
    & = e^{-
      a^{c}}\left\{\vphantom{\sum_{n}}G^{b}_{0}(E_{1})
    G^{c}_{0}(E_{2})G^{b\dagger}_{0}(E_{1})\right. \nonumber \\
&\hphantom{e^{-a^{c}}}+G^{b}_{0}(E_{1})\beta^{c}_{n}
    G^{c}_{n}(E_{2})\alpha^{c}_{n}G^{b\dagger}_{0}(E_{1}) \nonumber \\
    &\hphantom{e^{-a^{c}}}+ \beta^{b}_{n}G^{b}_{n}(E_{1})
    G^{c}_{n}(E_{2})\left.[\beta^{b}_{n}G^{b}_{n}(E_{1})]^{\dagger}
    \vphantom{\sum_{n}}\right\},
\end{align}
where $G^{i}_{0} = \langle \Phi^{i}_{0}|G^{i}|\Phi^{i}_{0}\rangle$,
and $G^{i}_{n} = \langle \Phi^{i}_{n}|G^{i}|\Phi^{i}_{n}\rangle$.
This result can now be written in terms of a double convolution with
a spectral function, i.e., 
\begin{align}
  \langle \Phi_{0}&|G^{b}(E_{1})
    G^{c}(E_{2})G^{b\dagger}(E_{1})|\Phi_{0}\rangle  \nonumber \\
    & = \int d\omega_{1}d\omega_{2}~
    A_{\rm eff}(E_{1},E_{2},\omega_{1},\omega_{2})  \nonumber \\
    &\times G^{b}_{0}(E_{1} -
    \omega_{1})
	  G^{c}_{0}(E_{2} - \omega_{2})G^{b\dagger}_{0}(E_{1} - \omega_{1}),
\end{align}
where the effective spectral function is given by
\begin{align}
    &A_{\rm eff}(E_{1},E_{2},\omega_{1},\omega_{2})  \nonumber \\
    &= e^{-a^{c}}
    \left\{\vphantom{\sum_{n}}\right.\delta(\omega_{1})\delta(\omega_{2}) \nonumber \\
      &+ \sum_{n}\left[
     \beta^{c}_{n}(E_{2})
     \alpha^{c}_{n}(E_{2})\delta(\omega_{1})\delta(\omega_{2} - \omega_{n})
     \right.\nonumber\\ 
     & + \left.\left.|\beta^{b}_{n}(E_{1})|^{2}\delta(\omega_{1} -
     \omega_{n})\delta(\omega_{2} - \omega_{n})\vphantom{\beta^{c}_{n}} 
\right]\vphantom{\sum_{n}}\right\}.
\end{align}

\end{document}